\documentclass[aps,prl,twocolumn,showpacs,superscriptaddress]{revtex4}

\usepackage{graphicx}

\begin{document}

\title{Direct observation of band-gap closure for a semiconducting carbon nanotube in a large parallel magnetic field}

\author{S. H. Jhang}\altaffiliation{e-mail:sungho.jhang@physik.uni-regensburg.de}
\affiliation{Institute of Experimental and Applied Physics,
University of Regensburg, 93040 Regensburg, Germany}

\author{M. Marga\'{n}ska}
\affiliation{Institute for Theoretical Physics, University of
Regensburg, 93040 Regensburg, Germany}

\author{Y. Skourski}
\affiliation{Dresden High Magnetic Field Laboratory,
Forschungszentrum Dresden-Rossendorf, 01328 Dresden, Germany}

\author{D. Preusche}
\affiliation{Institute of Experimental and Applied Physics,
University of Regensburg, 93040 Regensburg, Germany}

\author{M. Grifoni}
\affiliation{Institute for Theoretical Physics, University of
Regensburg, 93040 Regensburg, Germany}

\author{J. Wosnitza}
\affiliation{Dresden High Magnetic Field Laboratory,
Forschungszentrum Dresden-Rossendorf, 01328 Dresden, Germany}

\author{C. Strunk}
\affiliation{Institute of Experimental and Applied Physics,
University of Regensburg, 93040 Regensburg, Germany}
\date{\today}

\begin{abstract}
We have investigated the magnetoconductance of semiconducting carbon nanotubes (CNTs) in pulsed, parallel
magnetic fields up to 60~T, and report the direct observation of the predicted band-gap closure and the reopening of the gap under variation of the applied magnetic field.
We also highlight the important influence of mechanical strain on the magnetoconductance of the CNTs.
\end{abstract}


\maketitle


Carbon nanotubes (CNTs) are attractive building blocks for nanoelectronic devices.
While electronic properties of CNTs are determined to be either metallic or semiconducting once they are grown,
a magnetic field $B_\parallel$ parallel to the tube axis provides an elegant way to tune the band structure of a CNT after its growth \cite{Ajiki}.
The origin of the sensitivity to $B_\parallel$ lies in the contribution of the Aharonov-Bohm (AB) phase to the orbital phases picked
up by electrons encircling the perimeter of the tube.
The AB phase tunes the periodic boundary condition along the tube circumference
and results in a $\phi_0$-periodic modulation of the band gap \cite{Ajiki,Coskun,Lassagne}, where $\phi_0=h/e$ is the flux quantum.
Recently, significant drops in conductance $G$ were induced by $B_\parallel$ for initially metallic CNTs \cite{Lassagne,Fedorov2,Jhang}
as the energy gap of metallic CNTs linearly opens with magnetic flux $\phi$ for $\phi$ $\leq$ $\phi_0$/2.
For semiconducting CNTs, theory predicts that
the initial energy gap linearly decreases with $\phi$ to close the gap at $\phi=\phi_0$/3,
and reopens reaching a local maximum at $\phi=\phi_0$/2.
The gap then closes again at $\phi=2\phi_0$/3 and recovers its original value at $\phi=\phi_0$ \cite{Ajiki,Charlier}.
However, as actual magnetic fields $B_{\text{0}}$ equivalent to $\phi_0$ are about 5000 and 50~T for CNTs with diameters $d$ of 1~and 10~nm, respectively,
the AB effect of semiconducting CNTs has only been partially investigated for $\phi \ll \phi_0$ \cite{Minot,Zaric,Fedorov3},
and the direct observation of the predicted semiconductor-to-metal transition at $\phi=\phi_0$/3 has so far remained elusive.
Moreover, while CNTs of $d\ge$~5.5~nm are necessary to achieve $\phi_0$/3 within the accessible fields of about 60~T in a specialized pulsed-magnet lab,
the magnetoconductance (MC) in thick CNTs is often strongly affected by disorder and quantum interference effects \cite{Stojetz,Fedorov},
making it difficult to solely identify the AB effects on the band structure.

In this Letter, we report a magneto-transport study on a clean semiconducting CNT performed in pulsed magnetic
fields of up to 60~T.
The MC of the tube showed a clear manifestation of the AB effect on the band structure when located near the charge neutrality point (CNP).
The conductance changes with $B_\parallel$ by more than 100 times showing a peak, then a dip close to $B_\parallel=B_{\text{0}}$/2 before approaching the second peak.
The position of the peak is shifted from the expected $B_\parallel=B_{\text{0}}$/3,
which can be explained by the effect of mechanical strain originating from the tube bending.

\begin{figure}[b]
\includegraphics[width=8.2cm]{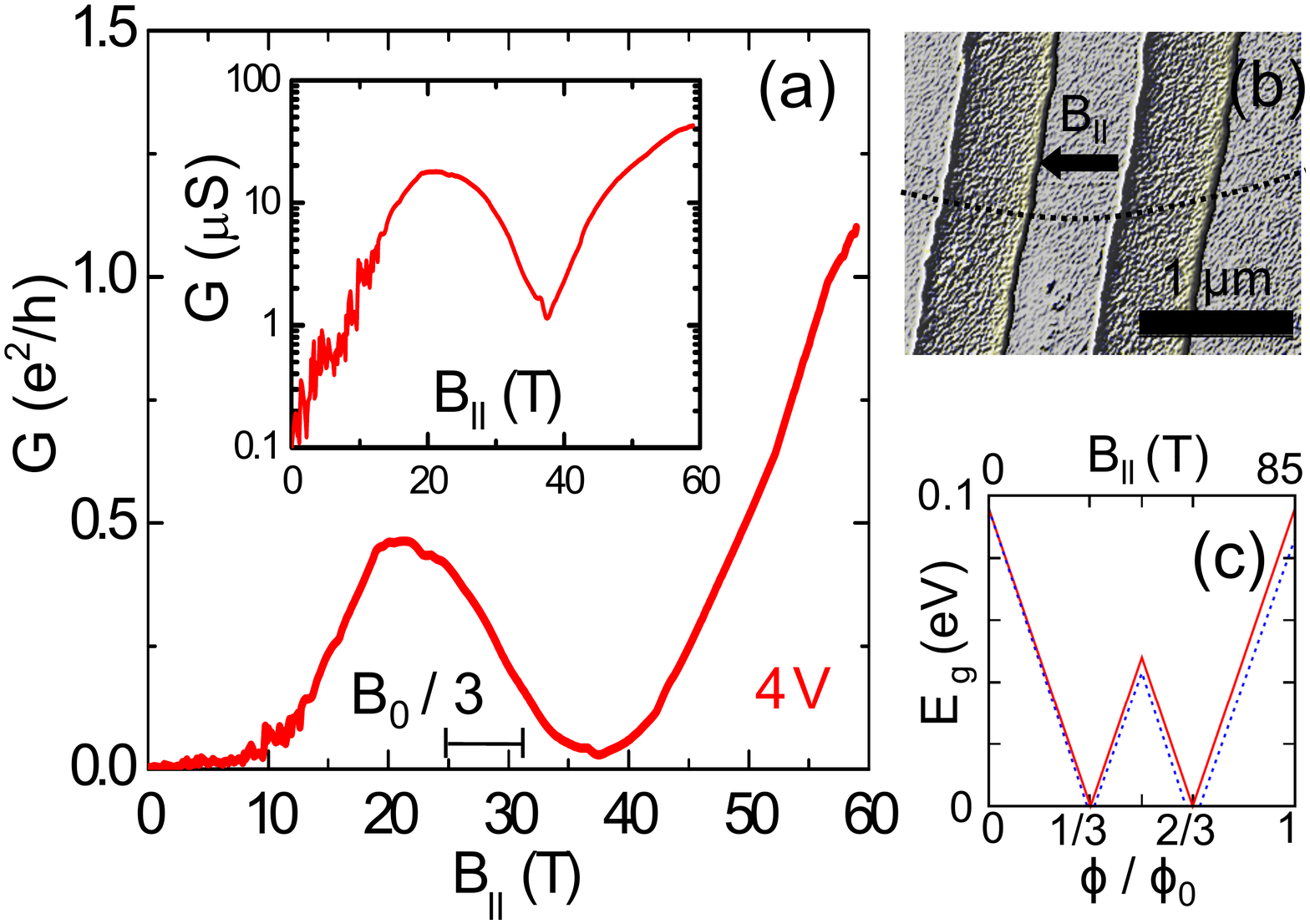}
\caption{\label{fig:epsart}
(a) MC of a semiconducting CNT device near the charge neutrality point, measured at 3.1 K.
The inset shows the MC in a semilog scale.
(b) The scanning electron microscope image of the measured device. The tube was smoothly bent while growing over 10 $\mu$m on the substrates
(black dotted line below the tube given as a guide line to the eyes).
The MC was studied between the two Pd electrodes, where the tube is almost linear.
Magnetic fields were applied parallel to this section of the tube, and the accuracy of the alignment was $\sim \pm 5^\circ$.
(c) Calculated energy gap of a (95,15) semiconducting CNT in a parallel magnetic field. Solid and dashed lines are without and with the Zeeman effect, respectively.
While the observed MC peak at $B_\parallel=$~22~T can be related to the gap closure at $\phi=\phi_0$/3, the diameter $d=8\pm0.5\;$nm, determined from the atomic force microscope,
rather suggests 24.5~$\leq B_{\text{0}}$/3~$\leq$~31.5~T (see the bar in (a)).
}\label{fig:MC}
\end{figure}

Our experiments have been performed on devices made of individual CVD-grown CNTs on Si/SiO$_{2}$/Si$_{3}$N$_{4}$ substrates \cite{Kong}.
The heavily $p$-doped Si was used as a back gate and the thickness of the insulating layer was 350~nm.
Pd (50~nm) electrodes were defined on top of the tubes by e-beam lithography. The distance between two Pd electrodes was $\sim$~500~nm. The dc two-probe
magnetoconductance was studied in pulsed magnetic fields, applied parallel to the tube axis.
The data presented here were obtained from a CNT ($d \approx$~8~nm) smoothly bent with a nearly constant curvature on the substrate as shown in Fig.~\ref{fig:MC}(b).

Fig.~\ref{fig:MC}(a) displays the MC trace of the semiconducting CNT, measured at 3.1~K.
The conductance at zero field $G$(0) is greatly suppressed as the Fermi energy is located near the CNP by applying a gate voltage $V_{g}= 4$~V.
With the application of $B_\parallel$, $G$($B_\parallel$) exponentially increases by two orders of magnitude to recover the level of one conductance quantum ($e^2/h$)
until it reaches a peak at $B_{\text{1}}=$~22~T. The conductance then drops back to a minimum around $B_{\text{min}}=$~37~T, before increasing
towards the expected second peak.
Although the second peak was not reachable in our experiment, the negative curvature of the MC curve near 60~T,
seen in the inset of Fig.~\ref{fig:MC}(a), indicates that the second peak is located not far above 60~T.
We also notice $G(0)\ll G$($B_{\text{min}}$).
For comparison, we calculated the energy gap $E_{g}$ in $B_\parallel$ for a (95,15) CNT (with $d$ similar to our tube) presented in Fig.~\ref{fig:MC}(c);
the large consecutive change in the conductance agrees in general with the band-gap modulation due to the AB effect.
The conductance peak at $B_{\text{1}}=$~22~T and the minimum at $B_{\text{min}}=$~37~T can be attributed to the band-gap closure at $\phi=\phi_0$/3
and a local $E_{g}$ maximum at $\phi=\phi_0$/2, respectively.
The observation, $G(0)\ll G$($B_{\text{min}}$), results from the fact that $E_{g}$($\phi_0$/2)~=~$\frac{1}{2}$~$E_{g}$(0).
However, we note two experimental observations not explained within the simple model:
1) The height of the first peak is smaller than that of the second peak.
2) If $B_{\text{1}}=$~22~T corresponds to the band-gap closure at $\phi=\phi_0$/3, then the second peak should already appear at $B_{\text{2}}=$~44~T.
Also, the diameter $d=8\pm0.5\;$nm, determined from the atomic force microscope (AFM), rather suggests 24.5~$\leq B_{\text{0}}$/3~$\leq$~31.5~T.

\begin{figure}[t]
\includegraphics[width=8.2cm]{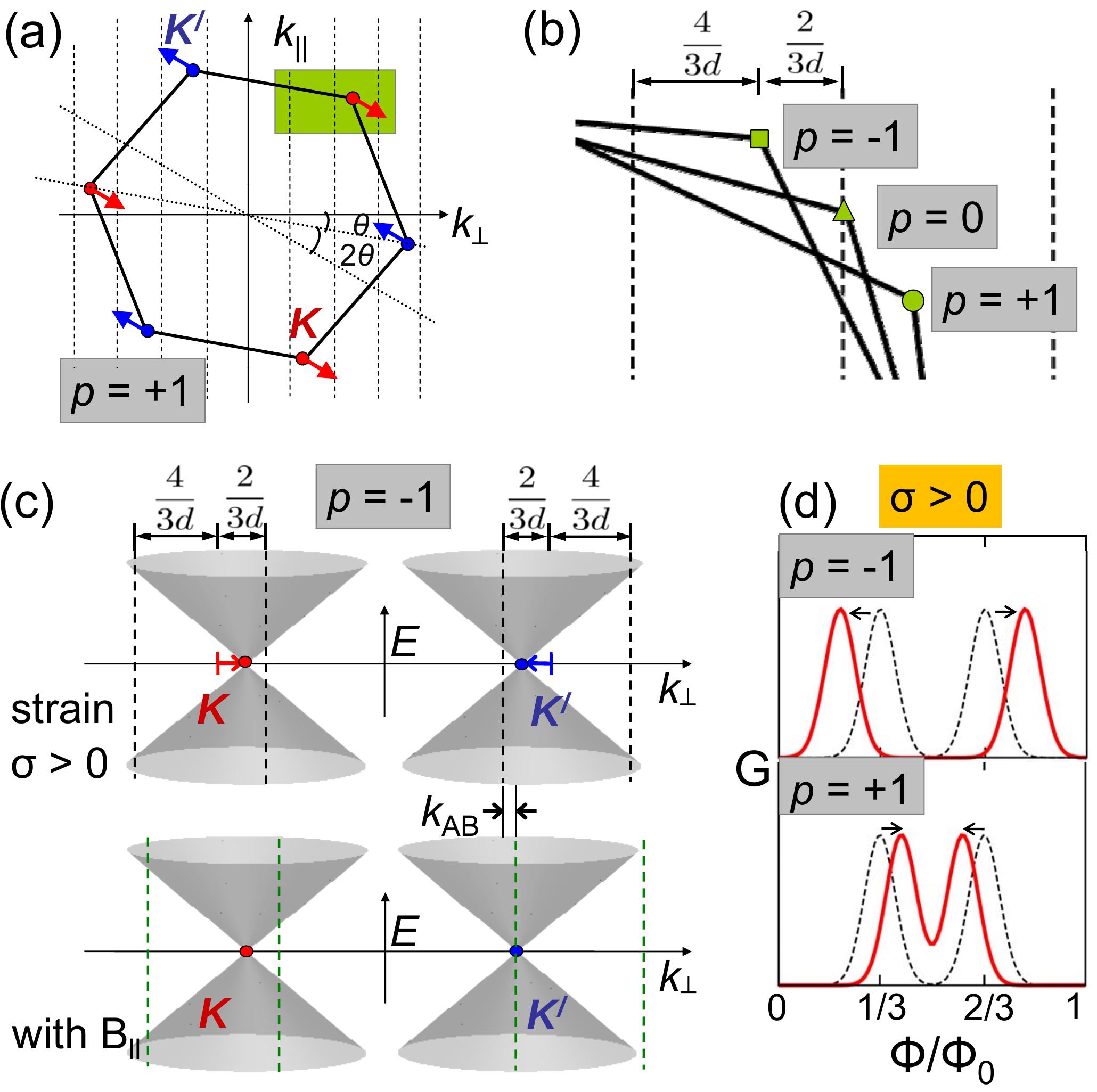}
\caption{
(a) Hexagonal Brillouin zone with lines of allowed $k_\perp$.
$K$(red) and $K'$(blue) points shift with an angle of 3$\theta$ from the $k_\perp$ axis in the presence of uniaxial strain.
(b) A zoom into the area bound by the green box shows the position of the $K$ point relative to the lines of allowed $k_\perp$, depending on the type $p$.
[Here $p=\pm1$ and $p=0$ are for semiconducting and metallic CNTs, respectively.]
(c) Corresponding Dirac cones with lines of allowed $k_\perp$ for the CNTs with $p=-1$ at $\sigma>$~0.
While upper cones display the shift of $K$ and $K'$ points under the strain,
lower cones explain the resulting effect with $B_\parallel$.
The allowed $k_\perp$ states shift to the right with increasing $B_\parallel$ by $k_{\text{AB}}=(2/d)(\phi_{\text{AB}} / \phi_0)$ due to the AB effect
and close the gap of the tube when crossing the $K$-points.
With the strain, those quantized $k_\perp$ lines intercept the $K'$ point earlier than at $\phi=\phi_0$/3, and the $K$ point later than at $\phi=2\phi_0$/3.
(d) Resulting shift of the MC peaks for CNTs at $\sigma>$~0.
Solid and dashed lines are with and without the strain, respectively.
The two peaks move either closer (for $p=+1$) or away from each other (for $p=-1$).
For simplicity, we neglect the change in the $k_\perp$ values due to the diameter shrinkage under the strain.
}\label{fig:Model}
\end{figure}

In the following, we show that the shifted positions of the MC peaks can be explained by the effect of mechanical strain in CNTs.
The structural deformation, such as the strain and tube bending,
plays an important role in the electronic structure of CNTs \cite{Yang,Chibotaru,Minot2,Huang}.
Previous works \cite{Yang,Minot2,Huang} showed that the band gap of the CNTs changes by $\sim \pm$100 meV per 1\% strain, due to the shift of the $K$ and $K'$ Dirac points under the strain.
This change of the band gap ($\Delta E_{g}$) is independent of diameter, whereas $E_{g} \propto d^{-1}$ for semiconducting CNTs.
Therefore, the effect of strain becomes more important for larger diameter tubes.

Fig.~\ref{fig:Model} illustrates the shift of the $K$ and $K'$ Dirac points under uniaxial strain, and the resulting effects on the positions of the MC peaks.
The shift of the $K$-points depends on the uniaxial strain $\sigma \,(= \frac {L-L_{0}}{L_{0}})$ and the chiral angle $\theta$, and is given by \cite{Yang}
\begin{eqnarray}\nonumber\label{eq:k}
&&\Delta k_\perp = \,\tau\, a_{\text{0}}^{-1}\, (1+\nu)\,\sigma \,\cos(3\theta),\;\\
&&\Delta k_\parallel = -\,\tau\, a_{\text{0}}^{-1}\, (1+\nu)\,\sigma \,\sin(3\theta),\;
\end{eqnarray}
where $\tau = \pm 1$ for the $K$ and $K'$ Dirac points, $a_{\text{0}}$ is the C-C bond length, and $\nu$ being the Poisson ratio.
It is displayed in Fig.~\ref{fig:Model} for the case of tensile strain ($\sigma>$~0).

Due to the shift of the $K$-points under $\sigma>$~0,
the positions of the gap closure at $\phi=\phi_0$/3 and at $\phi=2\phi_0$/3 are also shifted either closer (for $p=+1$) or away from each other (for $p=-1$)
depending on the type $p$ of the semiconducting CNT (Fig.~\ref{fig:Model}(d)). Here $p=\pm1$ such that the chiral indices $(n,m)$ satisfy $n-m = 3q+p$ with $q$ being an integer.
For the case of compressive strain ($\sigma<$~0), the effects are opposite with the type $p$.
Supposing the type of our tube as $p=-1$ and $\sigma>$~0 (or $p=+1$ and $\sigma<$~0), we can explain the positions of the MC peaks in our data.
For tubes with $p=-1$ and $\sigma>$~0, the first peak at $B_\parallel = B_0$/3 is shifted to the left, and the second peak at $B_\parallel = 2B_0$/3 to the right by the amount of $\Delta B = (d/2)\, B_0 \, \vert \Delta k_\perp \vert$.
Assuming the shifted peaks at $B_{\text{1}}=$~22~T and $B_{\text{2}}\sim$~60~T, a simple calculation \cite{calculation} leads to $B_{\text{0}}\approx$~82~T and $\Delta B\approx$~5~T.
This value of $\Delta B$ corresponds to $\sigma \,\cos(3\theta)\approx$ 1.8 $\cdot$ 10$^{-3}$ from Eq.~(\ref{eq:k}), supposing $\nu\approx$~0.2 \cite{Sanchez}.
Therefore, even small axial strain $\sigma = 1.8 \cdot 10^{-3}$ for zigzag tubes ($\theta = 0 ^\circ$) would explain the shift of MC peaks observed in our data \cite{equation}.
Without the strain-induced shift, the model suggests the MC peaks occur at~27 and 55~T with the $\phi_0$-periodicity of~82~T.
This value of $B_{\text{0}}$, equivalent to $d = 8.1$~nm, now shows a good agreement with the diameter of our tube obtained from AFM.
Note that the diameter shrinkage due to the strain ($\Delta d = d \,\sigma \,\nu \approx 3\cdot 10^{-3}$~nm) is negligible.
The existence of a small strain is likely in our device,
as our tube was mechanically deformed during the growth on the SiO$_{2}$ substrates [Fig.~\ref{fig:MC}(b)] \cite{thermal}.

\begin{figure}[t]
\includegraphics[width=8.2cm]{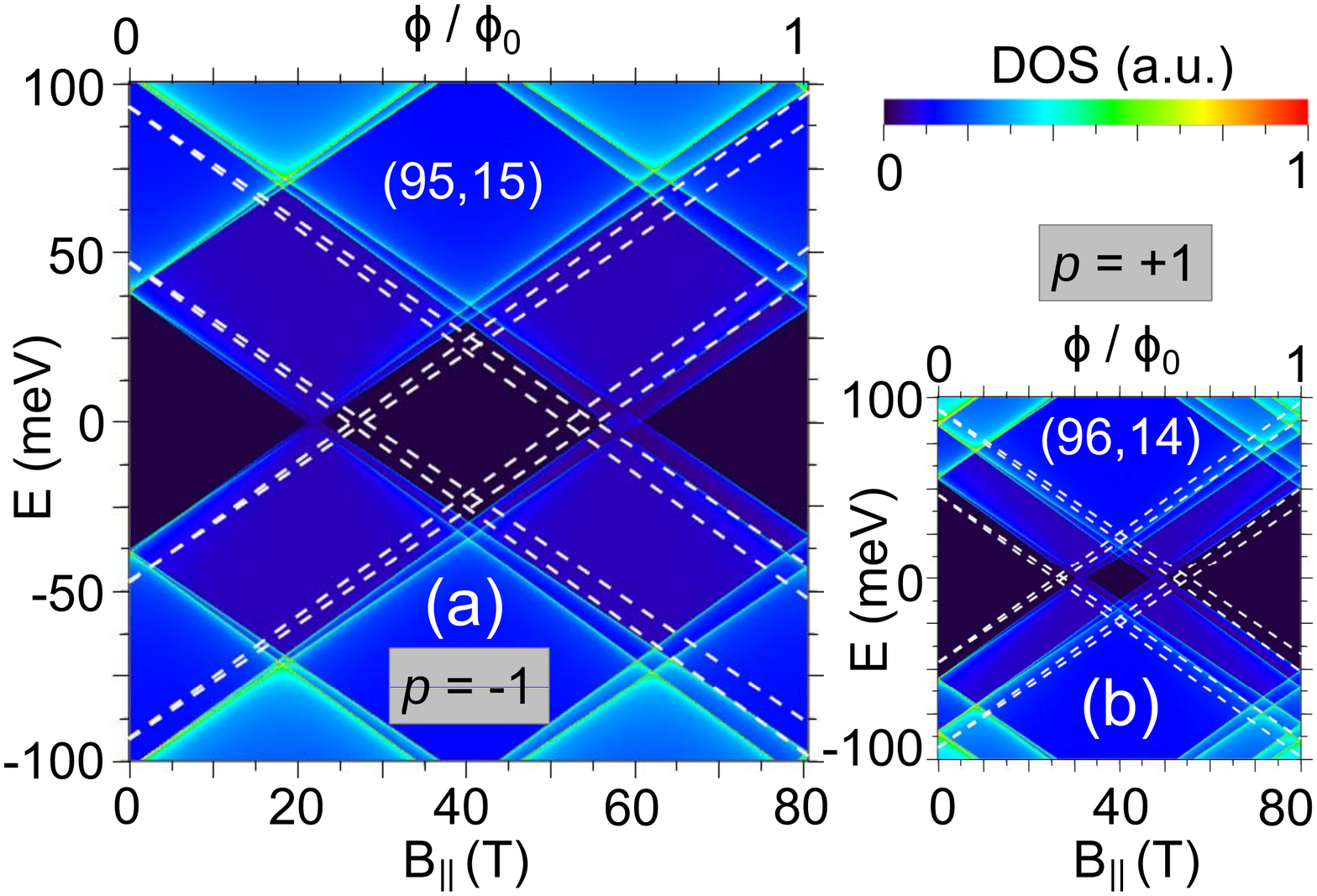}
\caption{
Model calculation of the DOS for (a) (95,15) and (b) (96,14) semiconducting CNTs in a parallel magnetic field at $+0.2\%$ strain.
For comparison, dashed lines indicate the band-edge position without strain.
}\label{fig:DOS}
\end{figure}

Taking the (95,15) tube ($d =$~8.1~nm and $\theta =7.2^\circ$) as a model CNT with $p=-1$, we calculated the density of states (DOS) in a parallel magnetic field at $+0.2\%$ strain.
For comparison, we present the DOS of a (96,14) tube, which has almost the same $d$ and $\theta$, but with $p=+1$.
The DOS was calculated from the dispersion relation, with momenta close to the Fermi points modified according to Eq.~(\ref{eq:k}).
We used periodic boundary conditions in the axial direction, suitable for very long nanotubes.
Displayed in Fig.~\ref{fig:DOS}, the dashed lines indicate the positions of the band edges without strain.
At zero field, the axial strain either reduces (for $p=-1$) or increases (for $p=+1$) the band gap of the CNTs,
as demonstrated by previous experiments \cite{Minot2,Huang}.
With the application of $B_\parallel$, the band edges evolve, reflecting the orbital and Zeeman splitting.
While the band gap is closed for both tubes at 27 and 54~T without strain,
the positions of the band-gap closure shift under $+0.2\%$ strain, resulting in the gap closure at 22 and 58~T for the (95,15) tube, and at 32 and 48~T for the (96,14) CNT.
The relation $E_{g} (\phi_0/2)=\frac{1}{2}$~$E_{g}$(0) becomes under strain
$E_{g} (\phi_0/2)>\frac{1}{2}$~$E_{g}$(0) (for $p=-1$) [or $E_{g} (\phi_0/2)<\frac{1}{2}$~$E_{g}$(0) (for $p=+1$)].

The DOS calculated for the (95,15) tube at $+0.2\%$ strain shows in general good agreement with the positions of the MC peaks for our tube.
However, the calculation neglects the tube bending and the coupling between different shells \cite{Kwon,Marganska} of multi-walled CNTs,
assuming the charge transport mainly through the outermost shell.
Also, quantum interference effects in the Fabry-Perot regime, such as the AB beating effect \cite{Cao} are ignored.
Therefore, we cannot expect to explain all features of the measured MC within our simple model.

If the charge transport also occurs through the inner shell,
the second peak at $B_{\text{2}}\sim$~60~T can be due to the band-gap closure (at $\phi=\phi_0$/3) from the inner shell,
while the first peak at $B_{\text{1}}=$~22~T originates from the outer shell of the tube.
However, corresponding $d$ of 5.4 and 9~nm for the inner and the outer shell, calculated from this model,
differ significantly from the known inter-shell distance in multi-walled CNTs~($\sim$~0.34~nm) \cite{Iijima}.

\begin{figure}[b]
\includegraphics[width=8.2cm]{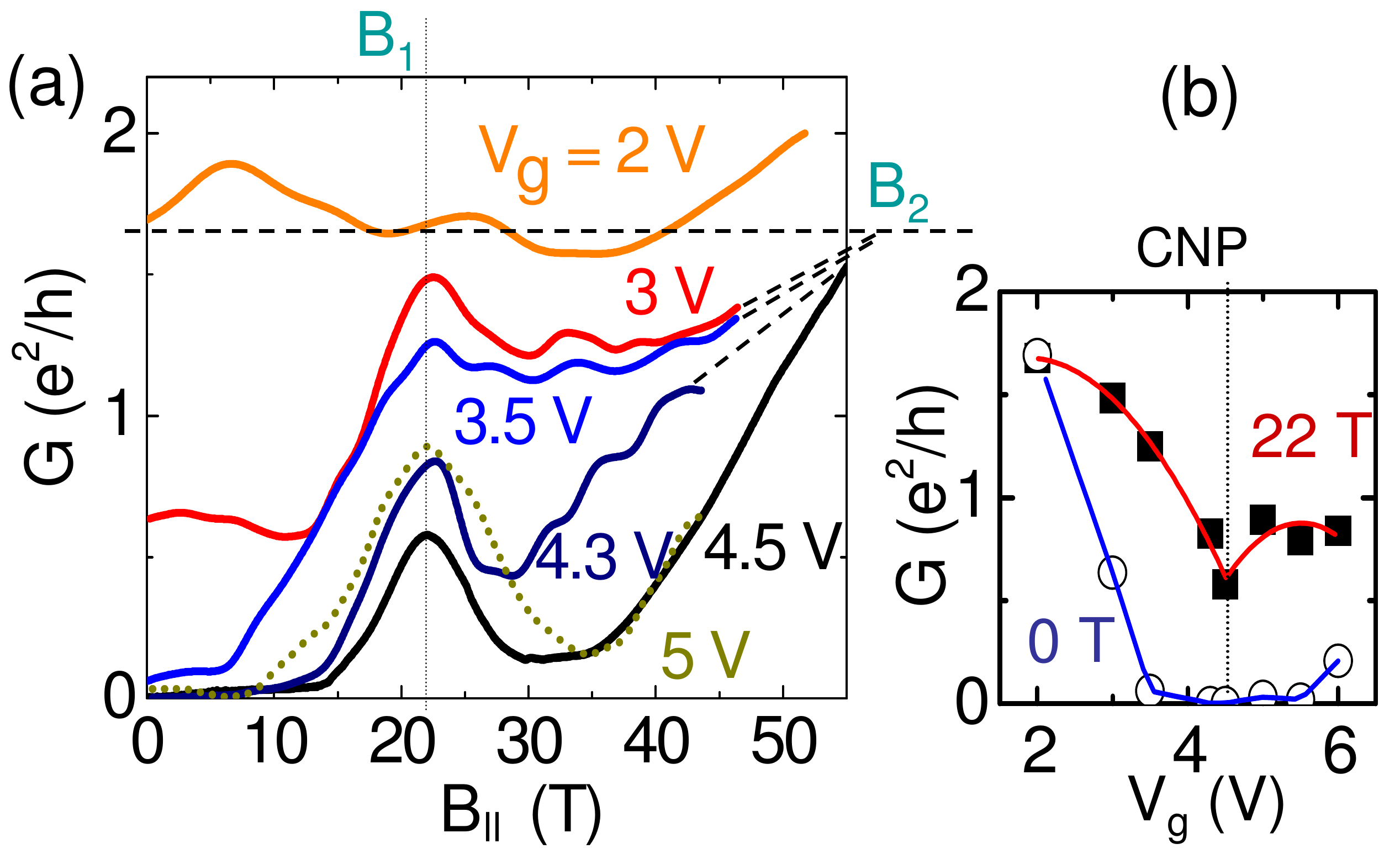}
\caption{\label{fig:epsart}
(a) $G(B_\parallel)$ traces at 3.5~K for various values of $V_{g}$ shown for the hole side of the CNP, except for $V_{g}=$~5~V (dotted line) on the electron side.
Near the CNP ($V_{g}^*\approx$~4.5~V), the $G(B_\parallel)$ data exhibit a peak at $B_{\text{1}}=$~22~T.
For $B_\parallel \geq$~45~T, the extrapolated curves seem to merge at one point,
implying the complete gap closure at $B_{\text{2}}\sim$~60~T (see the extrapolated dashed lines).
(b) The gate characteristics $G(V_{g})$ at $B_\parallel$~=~0 and 22~T, deduced from the $G(B_\parallel)$ traces.
Data points at $V_{g}=$~5.5 and 6~V are added from MC traces not shown in Fig.~\ref{fig:Gate}(a).
$G(V_{g})$ indicates that a small gap remains at $B_{\text{1}}=$~22~T.
Note the conductance for the electron side of the CNP is lower, compared to that of the hole side, due to the larger Schottky barriers at the CNT/Pd interface.
Solid lines are guides to the eyes.
}\label{fig:Gate}
\end{figure}

In order to confirm our interpretation of the data, we investigated the evolution of the
$G(B_\parallel)$ versus $V_{g}$.
In Fig.~\ref{fig:Gate}(a), MC traces at 3.5~K are displayed mainly for the hole side of the CNP, where the CNT/Pd interface is most transparent \cite{Interface}.
At $B_\parallel =$~0, the hole conductance at $V_{g} = $~2~V is high, almost 2$e^{2}/h$, and decreases rapidly
as the Fermi energy is tuned towards the CNP ($V_{g}^*\sim$~4.5~V).
Whereas the MC traces for $V_{g} \leq$~2~V stay flat except for small fluctuations,
a large conductance modulation appears only close to the CNP.

Fig.~\ref{fig:Gate}(b) presents the gate characteristics $G(V_{g})$ at 0 and 22~T, deduced from the MC traces in Fig.~\ref{fig:Gate}(a).
$G(V_{g})$ at $B_{\text{1}}=$~22~T shows that a small gap $\Delta_{\text{gap}}$ still remains at $B_{\text{1}}$,
whereas the AB effect predicts a complete gap closure at $\phi=\phi_0$/3.
On the other hand, the extrapolated MC curves converge at $B_\parallel \geq$~45~T [Fig.~\ref{fig:Gate}(a)], indicating a complete gap closure for the assumed second peak around 60~T.
The remaining small gap at $B_{\text{1}}$ is responsible for the smaller height of the first peak in Fig.~\ref{fig:MC}(a), compared to that of the second peak.

The Zeeman effect splits antiparallel spin states and reduces the band gap by $\Delta E_{\text{Zeeman}} \approx$~0.1~[meV/T]~$\cdot \, B_\parallel$,
affecting the $\phi_0$-periodic modulation of the band gap \cite{Coskun,Jiang} as shown in Fig.~\ref{fig:MC}(c).
Including the Zeeman effect, the small gap observed at $B_{\text{1}}$ is closed when the Zeeman contribution becomes larger than $\Delta_{\text{gap}}$ at higher $B_\parallel$.
The complete gap closure for the assumed second peak around 60~T suggests the size of $\Delta_{\text{gap}} <$~6~meV,
since $\Delta E_{\text{Zeeman}}$ corresponds to $\sim$~6~meV at 60~T.

Turning our attention to the origin of $\Delta_{\text{gap}}$ at $B_{\text{1}}$,
the tube bending can mix the states between the quantized lines of allowed $k_\perp$ and open a gap for metallic CNTs \cite{Chibotaru}.
Therefore, a bending-induced gap $\Delta_{\text{bend}}$, competing with $\Delta E_{\text{Zeeman}}$, is present at $\phi=\phi_0$/3 and at $\phi=2\phi_0$/3 for curved CNTs,
partly contributing to $\Delta_{\text{gap}}$.
However, $\Delta_{\text{bend}}\propto (d/D)^2$ \cite{Chibotaru}, with an estimated bending diameter $D$ of $\sim$~10~$\mu$m, is too small ($\ll$~1~meV) to explain the $\Delta_{\text{gap}}$.
On the other hand, the inter-shell interaction can also lead to a gap, for example, when the symmetry is lowered by disorienting one shell axis with respect to the other \cite{Kwon}.
Therefore, the observed $\Delta_{\text{gap}}$ at $B_{\text{1}}$ for our tube might originate from the inter-shell interaction, apart from the bending-induced gap.

Finally, we discuss the possible effect of spin-orbit coupling \cite{Kuemmeth,Ando} on the MC of semiconducting CNTs.
The spin splitting induced by spin-orbit coupling results in a peculiar double-peak MC structure for a chiral metallic CNT, as reported in our previous work \cite{Jhang}.
For semiconducting CNTs, the MC peak at $\phi=\phi_0$/3 does not split into two,
as the Zeeman contribution at $\phi=\phi_0$/3 ($\Delta E_{\text{Zeeman}}\approx 200/ d^{2}\,\text{meV}[\text{nm}^{-2}]$)
is much larger than the spin-orbit energy splitting ($\Delta_{\text{SO}}\approx 1.9/ d\,\text{meV}[\text{nm}^{-1}]$) \cite{SO}.

In conclusion, our experiment clearly shows that a semiconducting CNT can be converted into a metallic one with the application of large $B_\parallel$,
providing a consistent confirmation of the AB effect on the band structure of semiconducting CNTs.
In addition, we reveal that the position of the band-gap closure at $\phi=\phi_0$/3 can be tuned by mechanical strain.
Combined control of both the strain and the AB effect may open up new possibilities for magneto-electronic and magneto-optical CNT devices.

We acknowledge B. Witkamp and H. van der Zant for help in the growth of CNTs.
This research was supported by the Deutsche Forschungsgemeinschaft within GRK 1570 and SFB 689 and
by EuroMagNET under the EU contract No.\ 228043.


\begin{thebibliography}{}
\bibitem{equation} Eq.~(\ref{eq:k}) was confirmed by experiment \cite{Huang} with an additional prefactor of 0.57.
This would result in twice as large strain to explain the shift of the MC peaks in our data.
\bibitem{thermal} As our tube is fixed to electrodes on the substrate, different thermal-expansion coefficients between Si and CNT
can also play a role. Using $\alpha_{\text{Si}} \sim 2.6 \times 10^{-6}$~K$^{-1}$ and $\alpha_{\text{CNT}} \sim 10^{-5}$~K$^{-1}$ at room temperature \cite{Huang}, a rough estimate leads to $\sigma = (\alpha_{\text{Si}} - \alpha_{\text{CNT}}) \,\Delta T \approx 10^{-3}$,
with $\Delta T \approx -100$~K, consistent with our data.
\bibitem{Interface} Due to the difference in the work functions of the contact metal Pd and CNT, Schottky barriers at the CNT/metal interface become important for the electron side of the CNP. Therefore, the electron and the hole transmission probabilities are not symmetric with respect to the CNP owing to the difference in the contact properties; the conductance on the electron side is considerably lower as shown in Fig.\ 4(b). Hence, in the manuscript we focus on the hole side of the CNP, where we have the most transparent contacts at the CNT/Pd interface.
\bibitem{SO} For our tube with $d$ of 8~nm, $\Delta E_{\text{Zeeman}}\sim$~3~meV and $\Delta_{\text{SO}}\sim$~0.2~meV at $\phi=\phi_0$/3.
\end{thebibliography}
\end{document}